\documentclass[conference]{IEEEtran}
\IEEEoverridecommandlockouts
\usepackage{cite}
\usepackage{amsmath,amssymb,amsfonts}
\usepackage{algorithmic}
\usepackage{graphicx}
\usepackage{textcomp}
\usepackage{xcolor}
\usepackage[utf8]{inputenc}
\usepackage{booktabs}
\usepackage{hyperref}
\usepackage{array}
\usepackage{multirow}
\usepackage{caption}
\usepackage{subcaption}
\usepackage{placeins}

\makeatletter
\newcommand{\linebreakand}{%
  \end{@IEEEauthorhalign}
  \hfill\mbox{}\par
  \mbox{}\hfill\begin{@IEEEauthorhalign}
}

\renewcommand\subsubsection{\@startsection{subsubsection}{3}{\z@}%
  {1.5ex \@plus 1ex \@minus .2ex}%
  {0.5ex \@plus .2ex}%
  {\normalfont\normalsize\itshape}}
\makeatother

\def\BibTeX{{\rm B\kern-.05em{\sc i\kern-.025em b}\kern-.08em
    T\kern-.1667em\lower.7ex\hbox{E}\kern-.125emX}}

\title{Evaluating LLM Trade-offs for Enterprise Automation: Lessons from Workflow Generation in a Production Enterprise Platform}

\author{
  \IEEEauthorblockN{Xavier Wrenn} \IEEEauthorblockA{IBM\\ xavier.wrenn1@ibm.com} \and
  \IEEEauthorblockN{Radoslav Raykov} \IEEEauthorblockA{IBM\\ Radoslav.Raykov@ibm.com} \and
  \IEEEauthorblockN{Aleksandar Angelov} \IEEEauthorblockA{IBM\\ Aleksandar.Angelov1@ibm.com} \and
  \IEEEauthorblockN{Hirokuni Kitahara} \IEEEauthorblockA{IBM\\ hirokuni.kitahara1@ibm.com}
  \linebreakand
  \IEEEauthorblockN{Yuji Watanabe} \IEEEauthorblockA{IBM\\ muew@jp.ibm.com} \and
  \IEEEauthorblockN{Anca Sailer} \IEEEauthorblockA{IBM\\ ancas@us.ibm.com}
}

\begin{document}

\maketitle

\begin{abstract}
Enterprise compliance management requires rapid adaptation to evolving regulatory
frameworks (e.g., DORA, AI RMF, FedRAMP) and tight remediation SLAs. Traditional
static orchestrators often fail in hybrid cloud environments where event-driven
assessments triggered by configuration drift or incident response demand that
automation code adapt to runtime context in seconds. This paper presents lessons
learned from evaluating six large language models for AI-driven workflow generation
in a Production Enterprise Platform, benchmarked across 29 real-world IT automation
scenarios, two generation pipeline architectures, and eight independent runs per
prompt-model-pipeline configuration (2,784 runs total).

Our initial pipeline used monolithic workflow generation, achieving 31.5--82.8\%
structural success rates (JSON schema validity and correct UI rendering), with most
models struggling severely on complex JSON generation tasks. We developed a
redesigned piecewise generation pipeline that decomposes workflow construction into
variable scaffolding, base block assembly, and nested block generation. This
architectural shift increased structural success to 74.1--97.8\% across all models.

We analyze critical production tradeoffs, including cost (\$0.008--\$0.20/workflow),
latency ($<$50s for interactive use), and model selection. We analyze how the
piecewise decomposition enables cost-efficient, smaller models (e.g., mistral-small
at 95.7\% structural success and \$0.01/workflow) to reach production viability,
removing dependency on expensive frontier models. While mistral-medium-2505 and
gpt-oss-120b achieved the highest structural success rates (96.1\% and 97.8\%
respectively), mistral-medium-2505 carries a 19$\times$ cost premium versus
mistral-small. Our deployment lessons highlight the need to separate structural
validity from semantic correctness (logical fulfillment of user intent) and provide
a solution for model-agnostic, scalable automation in cloud engineering.
\end{abstract}

\begin{IEEEkeywords}
Large language models, workflow automation, AI-driven automation, Concert Workflows, performance evaluation, structural validation
\end{IEEEkeywords}

\section{Introduction}

IBM\textsuperscript{\textregistered} Concert Workflows (CW) is an advanced low-code
automation and orchestration tool that enables users to create network and IT
automation workflows. Rather than relying on traditional code-heavy scripting
approaches, CW simplifies automation by using discrete low-code action
blocks---unique actions applicable to hundreds of IT, networking, and cloud systems.
Users can drag and drop these building blocks to create powerful automations without
manual coding. This approach transforms API code into deployment-ready action blocks,
enabling organizations to rapidly develop or reuse existing automation scripts.

The introduction of CW AI advances this capability by providing a natural language
interface where users can describe the desired automation in plain language, and the
system automatically generates the corresponding low-code workflow. This feature
significantly reduces the complexity of building workflows and empowers rapid
low-code development. Key benefits include reduction in time to market for new
automations and services, accelerated mean time to resolution for IT and network
issues, and enhanced security and compliance through centralized policy control and
role-based access control.


This study evaluates the performance and accuracy of six large language models
for creating low-code workflows. We compare performance across two versions of CW AI
to benchmark improvements to the workflow generation system while controlling for
the underlying LLMs. Our primary evaluation metric is structural success, defined as
the generation of workflows that conform to CW's JSON schema, pass validation checks,
and render correctly in the UI. This includes valid action block references, correct
input and output parameter types, proper variable wiring, and syntactic correctness.

It is important to note that this metric does not assess whether workflows logically
fulfill user intent, which we term semantic correctness. A workflow can be
structurally sound while selecting incorrect action blocks or producing incomplete
automation. Semantic accuracy is addressed separately in Section~\ref{sec:semantic}
via an external LLM judge. In addition to structural and semantic correctness, we
track generation time (latency), token consumption (both input and output), and cost
per workflow to provide a comprehensive view of model performance across multiple
dimensions.

\section{Related Work}

Concert Workflows AI builds on several intersecting lines of research. The ReAct
framework~\cite{yao2023react} established the pattern of interleaving reasoning
traces and tool calls that underlies our pipeline's Reasoning agent; our v2
architecture externalizes the Planner step from that loop to reduce per-call
complexity. Retrieval-augmented generation (RAG)~\cite{lewis2020rag} provides the
theoretical basis for our block and workflow semantic search nodes, grounding
generation in schema-correct examples rather than relying on parametric knowledge
alone. Chain-of-thought prompting~\cite{wei2022cot} motivates our piecewise
decomposition: by decomposing workflow construction into variable scaffolding, base
blocks, and nested blocks, we reduce the cognitive load of each generation step in a
manner analogous to intermediate reasoning steps.

For API-grounded generation specifically, Gorilla~\cite{patil2024gorilla}
demonstrated that retrieval-aware training substantially reduces hallucination when
LLMs must select from large, overlapping API catalogs---a challenge directly
analogous to Concert Workflows' action block selection. On evaluation methodology,
HumanEval~\cite{chen2021codex} established the canonical pattern of measuring
functional (structural) correctness separately from semantic correctness, a
distinction we preserve explicitly: our Fact-checker measures JSON schema validity
and UI renderability, while semantic evaluation via a judge agent provides the
necessary additional validation~\cite{zheng2023judging}. The cost-accuracy tradeoff
we surface across the six models reflects concerns formalized in
FrugalGPT~\cite{chen2023frugalgpt}, which showed that model selection and cascade
strategies can match frontier model accuracy at significantly lower cost.
Grammar-constrained decoding~\cite{geng2023grammar} represents a complementary
approach to enforcing structural validity at inference time; our pipeline instead
applies post-hoc schema validation with iterative JSON Patch correction.


\section{Pipeline Evolution and Motivation}

The initial pipeline approach was to generate complete workflows in a single LLM
call. This monolithic generation strategy required the model to simultaneously
declare all workflow variables with correct types, instantiate top-level action
blocks with proper parameters, generate nested control structures such as loops and
conditionals, and wire inputs and outputs correctly across all blocks. For most
models, this cognitive load exceeded their structured output capacity, resulting in
malformed JSON, incorrect block references, and type mismatches.

Furthermore, the JSON Patch correction mechanism employed in the original pipeline
could fix simple syntactic errors but struggled with deeper structural problems like
incorrect variable declarations or malformed nested blocks. These limitations led us
to hypothesize that decomposing workflow generation into discrete, sequential stages
would reduce per-step complexity, allowing a broader range of models to succeed at
each stage independently.

\subsection{Architecture of Initial Pipeline (v1)}

The initial pipeline begins with a Classifier that interprets user prompts,
determines intent, and rewrites requests using explicit IT terminology. This
normalized prompt is then passed to a Reasoning module, which implements a
ReAct-style agent with retrieval-augmented generation (RAG) capabilities. The
reasoning agent orchestrates the workflow construction process by iteratively
analyzing the enriched user request, deciding which tools to call, and aggregating
retrieved context.

\subsubsection{Data Sources}

\textbf{Vendors} - list of all integration providers supported by the CW platform.
Each vendor corresponds to a set of action blocks that wrap that provider's APIs or
automation interfaces into reusable, low-code operations.

\textbf{1Punch} - collection of all available action blocks in the CW platform where
each entry is a JSON workflow made only of a single action block, and includes some
parameter typing built into the workflow. These are not end-user generated workflows
but rather standardized references that demonstrate the correct usage of a given
action block's required inputs, variables, and metadata.

\textbf{Blocks} - a dictionary that relates a vector embedding of an AI-generated
description (of an action block's purpose) to the unique internal ID of the action
block. Used to match a user's workflow request intent/prompt to a valid action block.

\textbf{Workflows} - analogous to the Blocks data source, each record pairs a vector
embedding (of an AI-generated description) to an example workflow.

\textbf{Processed integrations} - consists of all available action blocks and their
proper JSON structure. It is used for the ground truth reference when checking for
the correctness of a generated action block within a workflow.

\subsubsection{Nodes}

\textbf{Classifier} - serves as the first-stage LLM in CW, responsible for
interpreting raw user prompts. It determines the intent of the prompt and classifies
it as either reject, workflow generate, workflow edit, help, small talk. If not
rejected, it then rewrites the request into an enriched, structured command using
explicit IT terminology and embedded best-practice hints. This normalized prompt is
passed downstream to the Reasoning node.

\textbf{Reasoning} - implements a ReAct-style agent with RAG capabilities. It
orchestrates the workflow construction process by iteratively analyzing the enriched
user request, deciding which tools to call, and aggregating retrieved context. The
agent invokes retrieval tools such as Block search and Workflow semantic search until
sufficient information is gathered or a predefined iteration limit is reached.

\textbf{Block search} - a retrieval tool used by the Reasoning agent to identify
relevant integration blocks. For each inferred intent, the service performs parallel
similarity searches over block metadata, including 1Punch descriptions. Retrieved
blocks are ranked and returned to the agent as candidates for workflow construction.

\textbf{Workflow semantic search} - a retrieval tool that surfaces previously built
workflows that are conceptually similar to the current user request. It embeds the
raw user prompt, performs a similarity search over the Workflows data source, and
returns the top matches. These workflows are used as few-shot references during
generation to reinforce correct structure and reduce hallucinations.

\textbf{Planner} - a tool invoked by the Reasoning agent once sufficient context has
been collected. It expands the enriched user request into a structured, step-by-step
workflow plan. The plan describes the intended automation logic as an ordered
sequence of comments, actions, and variable assignments, abstracting away concrete
API details while clearly specifying data flow and dependencies.

\textbf{Generate workflow} - consumes the structured plan, selected integration
blocks, and retrieved example workflows to generate a complete JSON workflow
conforming to CW's internal schema. The LLM is guided by a detailed system prompt to
ensure correct structure, block usage, and variable wiring.

\textbf{Fact-checker} - validates the structural and semantic correctness of
generated workflows by inspecting their JSON definition. It programmatically verifies
that all referenced action blocks exist, performs schema validation using Zod, checks
for required metadata and variables and ensures block inputs conform to expected
types. The Processed integrations serve as the ground truth for checking the
integrity of a generated action block. Errors and warnings are directly mapped to
specific lines in the JSON, providing traceable diagnostics for quality assurance. If
there are any errors found by the fact-checker, the generated workflow is fed to the
Generate JSON Patch together with the found errors. The number of fixing iterations
is manually adjusted, but we default to a maximum of 2.

\textbf{Generate JSON Patch} - triggered when the Fact-checker detects errors. It
produces a minimal JSON Patch that corrects the identified issues. The patched
workflow is revalidated, forming a feedback loop that continues until the workflow
passes validation or a retry limit is reached. A JSON Patch approach was used to
minimize the risk of changes in otherwise correct parts of a workflow as well as to
reduce the time needed for generation.

\subsubsection{Overview}

The CW initial AI pipeline begins with a Classifier that interprets the user request
and routes it based on intent or rejects it in cases of HAP or copyright violation.
Approved requests move on to Reasoning, which is an implementation of a ReAct
pattern with RAG capabilities. The reasoning agent performs tool calls such as Block
search to retrieve the most relevant action blocks via similarity search and
LLM-based ranking, and Workflow semantic search which supplements the generation with
previously-built, conceptually similar workflows as few-shot examples to improve
structure and reduce hallucination.

Once the agent is satisfied with the collection, or reaches a given limit, it
performs a Planner tool call which takes the user request prompt and expands it into
a structured, step-by-step workflow plan. Each plan outlines the intended automation
logic using a sequence of comments, actions, and variable assignments using the
gathered data as reference. The Generate workflow module uses the plan, selected
blocks, and examples to produce a complete JSON workflow. This output is then
validated by the Fact-checker, which inspects schema integrity, block existence, and
input correctness. If errors are found, a Generate JSON Patch module is invoked in a
feedback loop to correct the workflow.

\subsection{Architecture of the Advanced Pipeline (v2)}

The v2 pipeline retains the Classifier and basic retrieval infrastructure but
introduces several critical architectural changes. Most importantly, the Planner is
pulled out of the reasoning loop and invoked deterministically early in the process
to produce a first-pass workflow plan before any tool-based reasoning occurs. This
early planning provides structure that guides subsequent retrieval and generation
steps.

The reasoning loop then focuses exclusively on gathering implementation evidence for
the plan through an expanded set of tools. Block Semantic Search identifies the most
appropriate action blocks as before, but is now supplemented as follows.

\subsubsection{Additional Data Sources}

\textbf{Chunked data} - a collection of workflow chunks (sequences of action blocks)
extracted from high quality, user generated workflows. The addition of this chunked
data provides the model with additional context on how specific actions might
interact with one another to accomplish a specific task. While this chunked data is
not a full substitute for workflow samples it allows for fewer full workflow samples
to be passed in the context window while maintaining a high level of awareness to the
expected structure with a lower token consumption.

\textbf{Human readables} - Consists of a collection of structured, human readable
documents on how to format or use a specific action block. This documentation is an
AI-generated derivative of the vendor's original API documentation for a given
action. It provides details on parameter types, required or optional, description of
what each parameter is, and in some cases examples of the parameters passed into the
API call.

The following nodes are used.

\subsubsection{Nodes}

\textbf{Planner} - functions similarly to the v1 Planner with a few key differences.
The most important change was pulling the planner out of the reasoning loop and into
its own deterministically called node. This allows for a simpler and faster reasoning
loop as well as a more consistent and traceable planner. The core functionality still
remains as in v1 Planner but with a more detailed outline on required variables and a
better structure for listing nested blocks.

\textbf{Chunked data search} - a retrieval tool used by the Reasoning agent to
identify relevant chunks. Retrieved chunks are ranked and returned to the agent as
candidates for workflow construction.

\textbf{Document search} - a retrieval tool used by the Reasoning agent to query
user uploaded documents (if any are present). These document can help the agent
follow naming conventions and specific methodologies when generating workflows.

\textbf{Update plan} - once sufficient context has been collected, the pipeline
performs a quick revision to reconcile the initial plan with the retrieved evidence,
adjusting the plan based on actual available blocks and proven patterns. This step is
crucial to ensure workflow generation proceeds with structural and syntactic
awareness. The critical innovation in v2 is the decomposition of workflow generation
into three distinct, sequential stages: Generate Workflow Variables, Generate
Workflow Base Blocks, and Generate Workflow Nested Blocks.

\textbf{Generate workflow variables} - extracts the workflow variables outlined in
the planner and generates a JSON array containing each variable, its name, type, and
default value. These variables are then added to a blank workflow.

\textbf{Generate base blocks} - extracts the top level workflow blocks from the
planner and using the generated variables and reference usage data collected by the
reasoning agent builds each block, assigns input parameters and finally adds them to
the workflow JSON.

\textbf{Generate nested blocks} - similar to Generate base blocks, it extracts
workflow blocks from the planner, this time with a focus on nested blocks (nested in
loops and conditional statements). Each block is build using the same approach and
added to its corresponding parent. Breaking up the block generation step like this
allows for much deeper and more complex flow structures while maintaining structural
coherency.

\textbf{Attempt programatic fix} - an additional step in v2 following the Fact
checker node, which attempts simple programmatic fixes with a low time complexity
before resorting to LLM-based correction. This step handles simple programmatic
fixes with low time complexity, such as poorly escaped content, malformed action
blocks, and simple syntactical errors, before invoking Generate JSON Patch for any
remaining issues.

\subsubsection{Overview}

The CW v2 AI pipeline begins with the Classifier, which classifies and normalizes
the user request or rejects it for policy violations. For approved requests, v2
deterministically invokes the Planner early to produce a first-pass workflow plan
before any tool-based reasoning occurs. The system then enters a Reasoning loop
(ReAct + RAG) whose primary goal is to gather implementation evidence for the plan
via tool calls: Block semantic search to identify the most appropriate action blocks,
Chunked data semantic search to retrieve proven multi-block interaction patterns,
Workflow semantic search to provide a small number of structurally correct few-shot
workflow examples, and Document search to incorporate user-provided conventions when
applicable. Once sufficient context is collected, the pipeline performs an Update
plan step to reconcile the initial plan with the retrieved evidence. Workflow
generation is then decomposed into three stages: Generate workflow variables builds
the workflow variable scaffold, Generate workflow base blocks constructs the
top-level blocks, and Generate workflow nested blocks fills in blocks inside
control-flow structures. The resulting JSON is validated by the Fact-checker against
schema and processed integration ground truth. If errors are found, the pipeline
first runs Attempt programmatic fix for low-cost repairs; remaining issues are
corrected via Generate JSON Patch in a feedback loop until the workflow passes
validation or the maximum attempt limit is reached.

The architectural differences between the two versions can be summarized as follows.
The initial generation strategy employed a single monolithic call, while v2 uses
piecewise generation across three stages. Planner invocation was embedded inside the
reasoning loop in the initial version but moved to a deterministic early call in v2.
Context sources expanded from just Blocks and Workflows to include Chunked Data,
Human-Readable documentation, and user-uploaded Documents. Most critically,
complexity per LLM call decreased dramatically, with the initial pipeline requiring
the model to generate complete workflows in one shot while the redesigned pipeline
breaks this into manageable single-component tasks.

By decomposing workflow generation into variable scaffolding, then base blocks, then
nested blocks, each individual task falls within smaller models' capability
threshold. The model no longer needs to hold the entire workflow structure in context
simultaneously, instead building it incrementally through a series of focused
generation steps.

\section{Method}

\subsection{Experimental Setup}

We evaluated six large language models, ranging from 24 billion to 400 billion total
parameters. The granite-4-h-small model uses a mixture-of-experts (MoE) architecture
with 32 billion total parameters but only 9 billion active during inference. The
llama-3-3-70b-instruct is a 70-billion parameter dense model, while
llama-4-maverick-17b-128e-instruct-fp8 employs an MoE architecture with 400 billion
total parameters, 17 billion active parameters, and 128 expert modules.

The mistral-small-3-1-24b-instruct-2503 is a 24-billion parameter dense model, while
mistral-medium-2505 is approximately 70 billion parameters. Finally, gpt-oss-120b
uses an MoE architecture with 117 billion total parameters and 5.1 billion active
parameters. Pricing varies significantly across these models, with input token costs
ranging from \$0.00006 to \$0.003 per thousand tokens and output costs from
\$0.00025 to \$0.01 per thousand tokens. All models support a context window of
131,072 tokens (Table~\ref{tab:models}).

\begin{table*}[!t]
\centering
\caption{Models Compared}
\label{tab:models}
\small
\begin{tabular}{@{}lcccc@{}}
\toprule
\textbf{Model name} & \textbf{Model size} & \textbf{Input price} & \textbf{Output price} & \textbf{Context window} \\
 & \textbf{(in billion parameters)} & \textbf{(USD/1,000 tokens)} & \textbf{(USD/1,000 tokens)} & \textbf{(input + output tokens)} \\
\midrule
granite-4-h-small & 32 (9 active) & \$0.00006 & \$0.00025 & 131,072 \\
llama-3-3-70b-instruct & 70 & \$0.00071 & \$0.00071 & 131,072 \\
llama-4-maverick-17b-128e-instruct-fp8 & 400 (17 active) & \$0.00035 & \$0.0014 & 131,072 \\
mistral-small-3-1-24b-instruct-2503 & 24 & \$0.0001 & \$0.0003 & 131,072 \\
mistral-medium-2505 & $\sim$70 & \$0.003 & \$0.01 & 131,072 \\
gpt-oss-120b & 117 (5.1 active) & \$0.00015 & \$0.0006 & 131,072 \\
\bottomrule
\end{tabular}
\end{table*}

We selected 29 real-world compliance IT automation prompts based on actual user
requests to the CW AI system, spanning security posture tasks such as finding and
closing AWS Security Groups on specific ports, cloud infrastructure provisioning
including VPC topology creation and EC2 snapshot management, configuration
management operations such as RHSA patching with Ansible, monitoring and
observability workflows for SevOne device management and Instana metrics collection. The full
prompt list is available in the project repository in support of Open Science practices.

Data was collected using a unified experimental setup based on a modified version of
the production pipeline that enables extensive metrics collection. We executed eight
independent runs per prompt-model-pipeline combination, yielding 2,784 total runs (8
runs $\times$ 29 prompts $\times$ 6 models $\times$ 2 pipeline versions). Each
output was assessed for duration (total time from request to validated workflow),
token consumption (separated into input and output tokens), calculated cost based on
token consumption and published pricing, workflow generation success (whether any
workflow was produced versus rejection or pipeline failure), and structural soundness
(JSON validity, schema compliance, and UI renderability). Aggregated per-cell metrics
in Tables~\ref{tab:v210}--\ref{tab:improvement} report the mean across these eight
runs and the fraction of runs that passed each binary criterion.

Compared to the single-run pilot reported previously, this evaluation executes eight
runs per prompt-model-pipeline cell, mitigating much of the stochastic variance
inherent to LLM-based generation. Reported rates are therefore aggregated over 232
attempts per model per pipeline (8 runs $\times$ 29 prompts). The benchmarking
infrastructure (including the vLLM-based inference stack for gpt-oss-120b) was held
constant across both pipeline versions in this experiment, so the v1 versus v2
comparison reflects pipeline changes rather than environmental drift. We do not yet
report confidence intervals or significance tests at the per-cell level; that
analysis, including variance characterization by prompt difficulty, is left to future
work.

\subsection{Limitations}

Reported rates are point estimates aggregated over 8 runs per cell; per-cell
confidence intervals and significance tests are not yet computed and are left to
future work. Structural success measures JSON schema validity and UI renderability
only---semantic correctness is addressed separately in Section~\ref{sec:semantic}
but not yet integrated into the pipeline as a blocking validation step. The
gpt-oss-120b vLLM infrastructure was held constant across both pipeline versions in
this experiment, removing the infrastructure confound present in the prior
single-run pilot; however, the stub-workflow failure mode identified in
Section~\ref{sec:semantic} suggests a model-specific sensitivity to the v2 prompt
structure that warrants targeted follow-up.

\section{Results}

Tables~\ref{tab:v210}--\ref{tab:improvement} summarize structural success rates,
generation time, cost, and token consumption across both pipeline versions. Under v1,
performance varied widely: gpt-oss-120b led at 82.8\% structural success while
granite-4-h-small and mistral-small fell below 35\%. Under v2, all models improved
substantially---structural success rates rose to 74.1--97.8\%, with four of six
models exceeding the 90\% production-viability threshold, compared to zero under v1.

Several cross-cutting patterns emerge (Table~\ref{tab:improvement}). Every model
gained structural success under v2, validating the piecewise generation architecture
across the model size spectrum. The primary cost of this gain is latency: v2 requires
more sequential LLM calls, driving input token consumption higher for most models and
increasing average generation times by 8--75 seconds. gpt-oss-120b is the sole
exception---latency fell 33 seconds between versions. llama-3-3-70b-instruct is the
sole model whose generation \emph{rate} declined (95.3\% $\to$ 81.5\%), and it
carries the worst latency (151.39s) and cost (\$0.0785) of any v2 model. No model in
this 8-run evaluation reaches 100\% structural success.

To understand the production implications of costs, consider 1,000 workflows per day
(365,000/year): mistral-small $\approx$\$3,700/yr, llama-4-maverick
$\approx$\$8,900/yr, gpt-oss-120b $\approx$\$6,000/yr, mistral-medium
$\approx$\$71,900/yr.

\begin{figure*}[!t]
\centering
\begin{subfigure}[t]{0.49\textwidth}
    \centering
    \includegraphics[width=\textwidth]{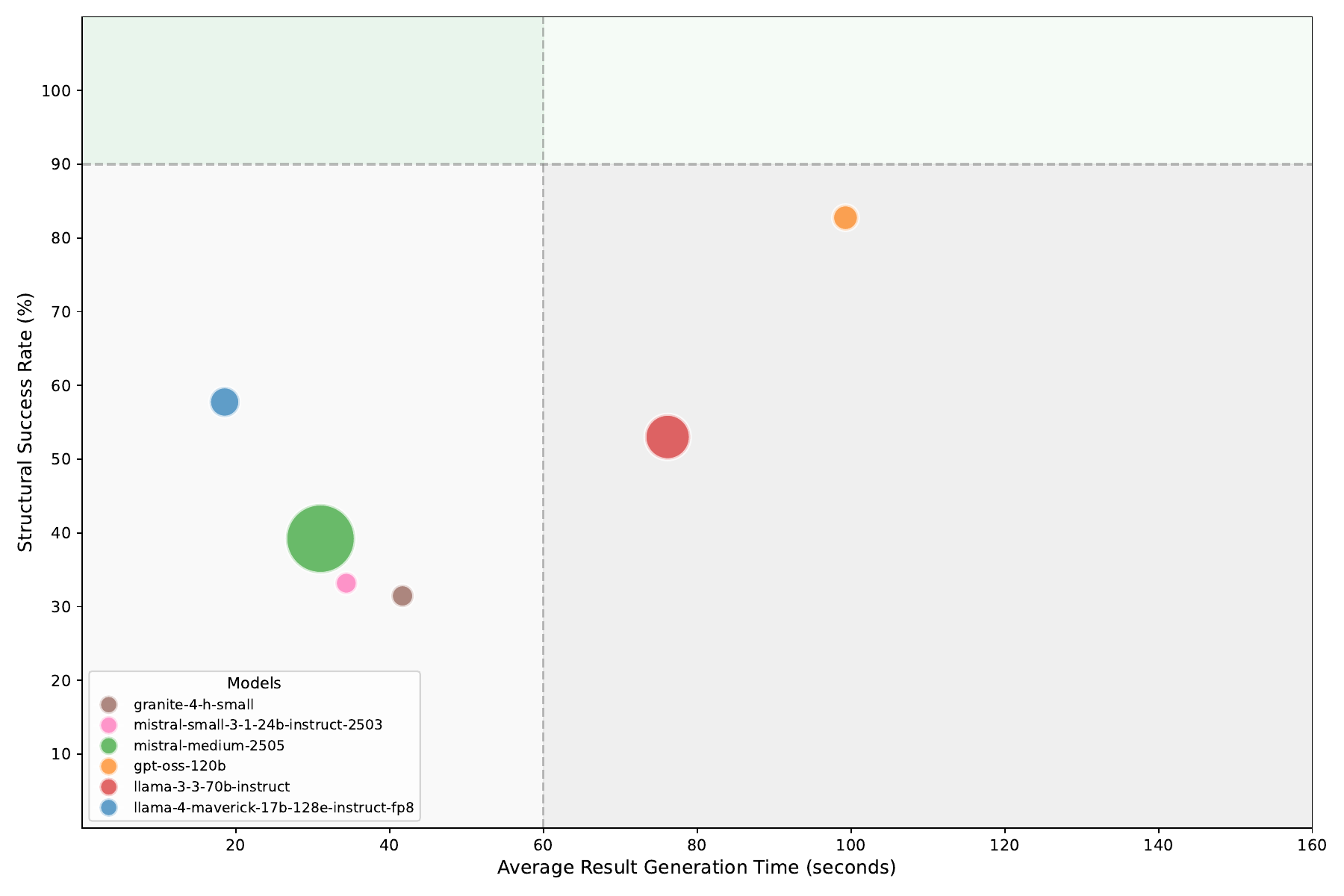}
    \caption{v1}
\end{subfigure}
\hfill
\begin{subfigure}[t]{0.49\textwidth}
    \centering
    \includegraphics[width=\textwidth]{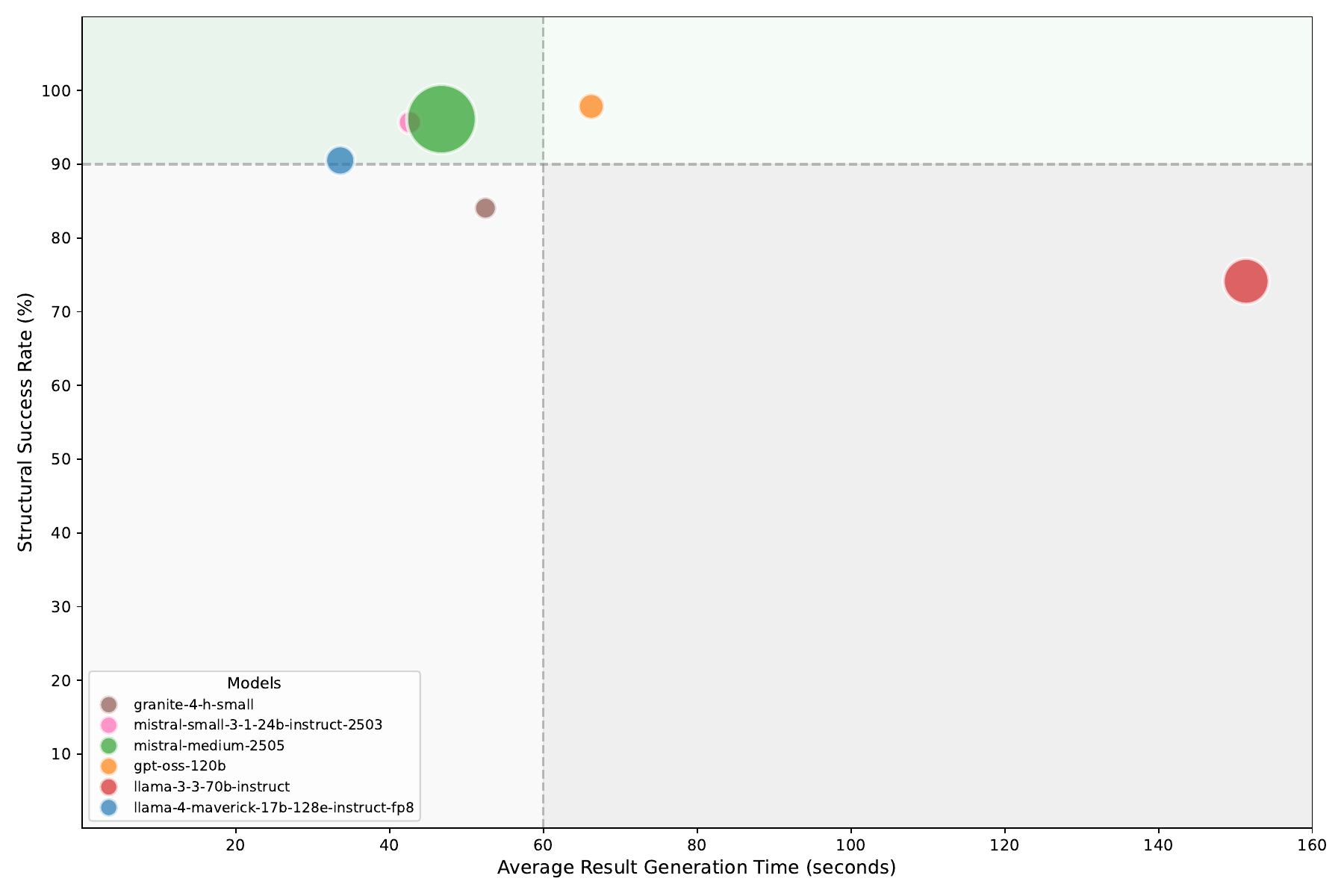}
    \caption{v2}
\end{subfigure}
\caption{Structural Success Rate (\%) vs Average Result Generation Time (seconds)
where larger circles represent greater token usage and cost.}
\label{fig:scatter}
\end{figure*}

\begin{figure*}[!t]
\centering
\begin{subfigure}[t]{0.49\textwidth}
    \centering
    \includegraphics[width=\textwidth]{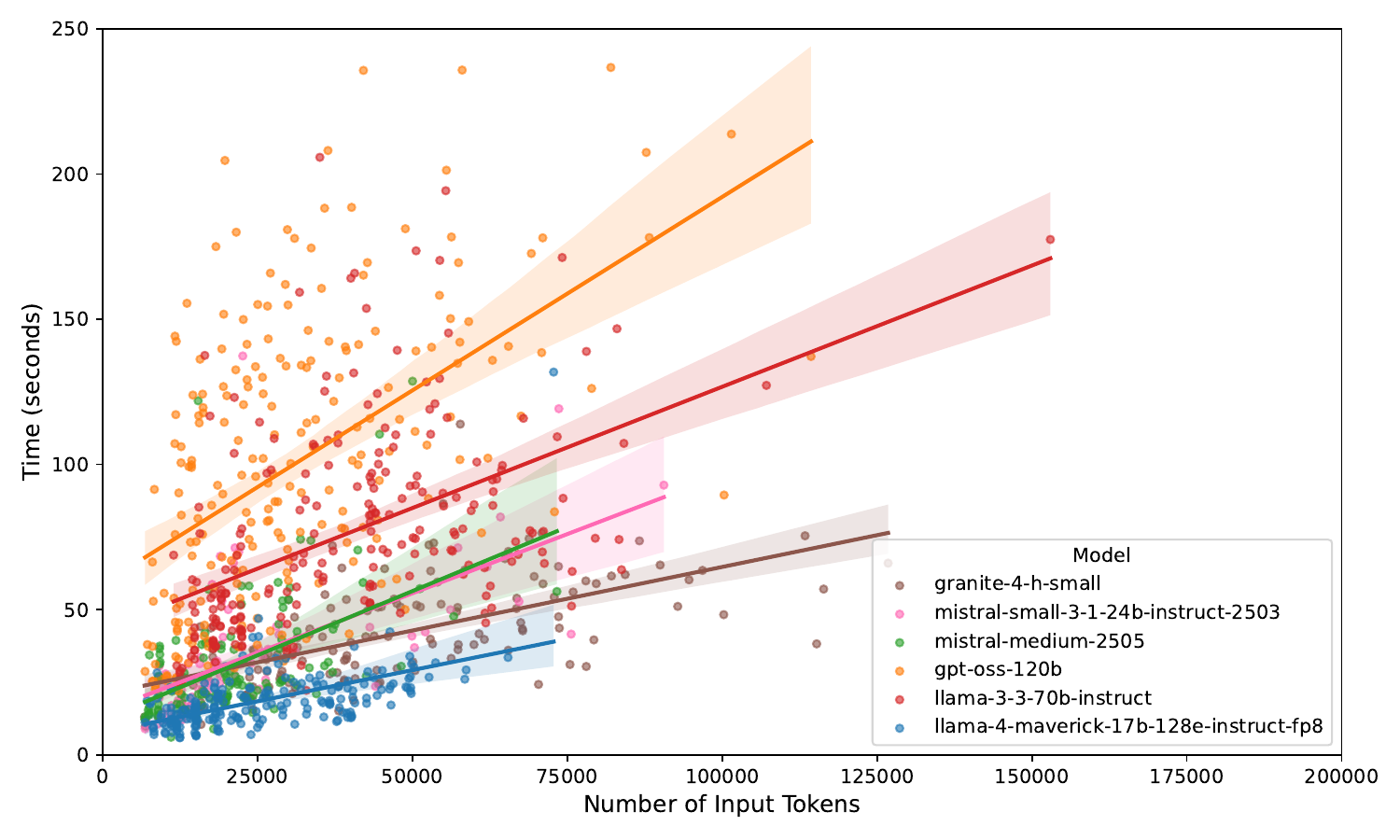}
    \caption{v1}
\end{subfigure}
\hfill
\begin{subfigure}[t]{0.49\textwidth}
    \centering
    \includegraphics[width=\textwidth]{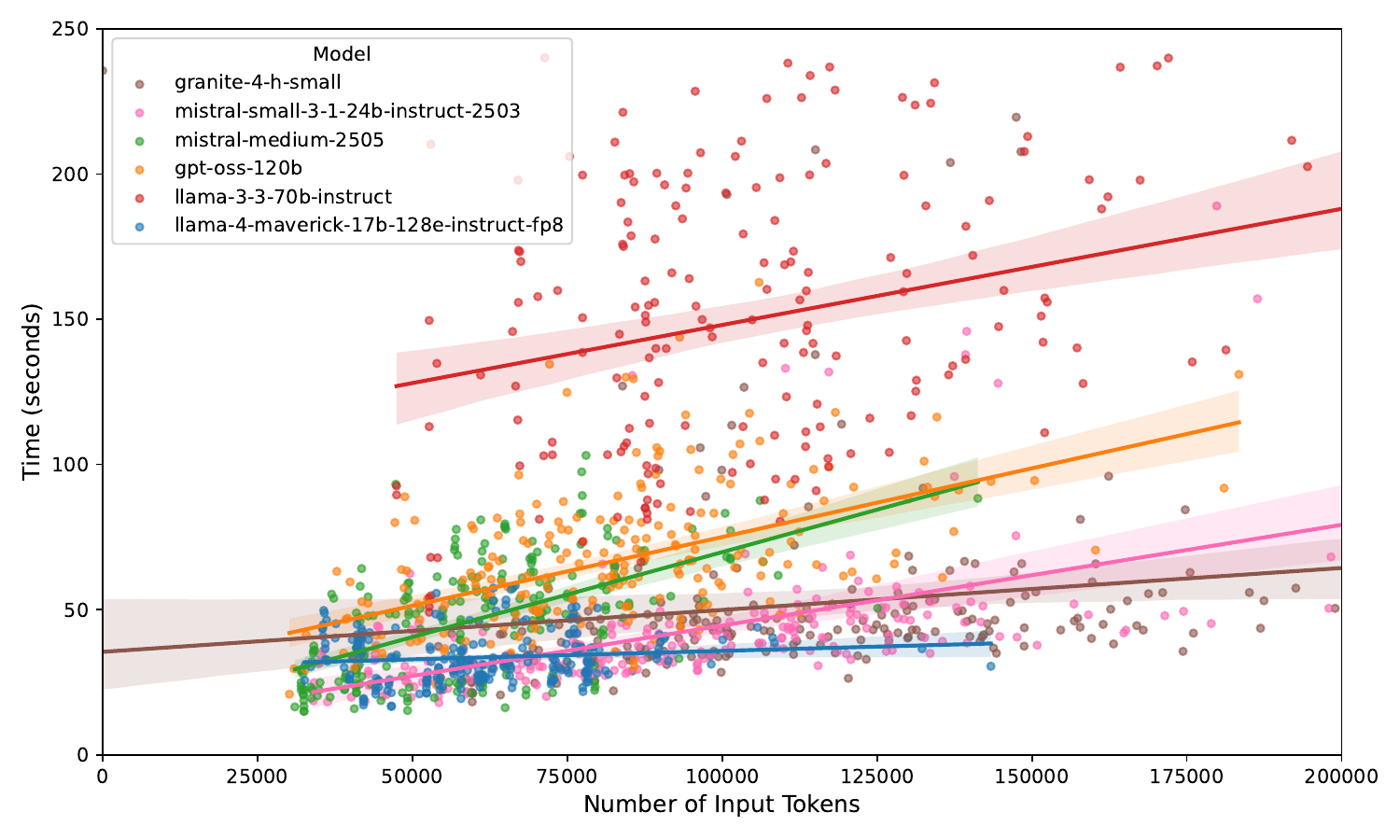}
    \caption{v2}
\end{subfigure}
\caption{Input tokens against time taken to generate a workflow of all successfully
generated workflows.}
\label{fig:tokens}
\end{figure*}

\begin{figure*}[!t]
\centering
\begin{subfigure}[t]{0.49\textwidth}
    \centering
    \includegraphics[width=\textwidth]{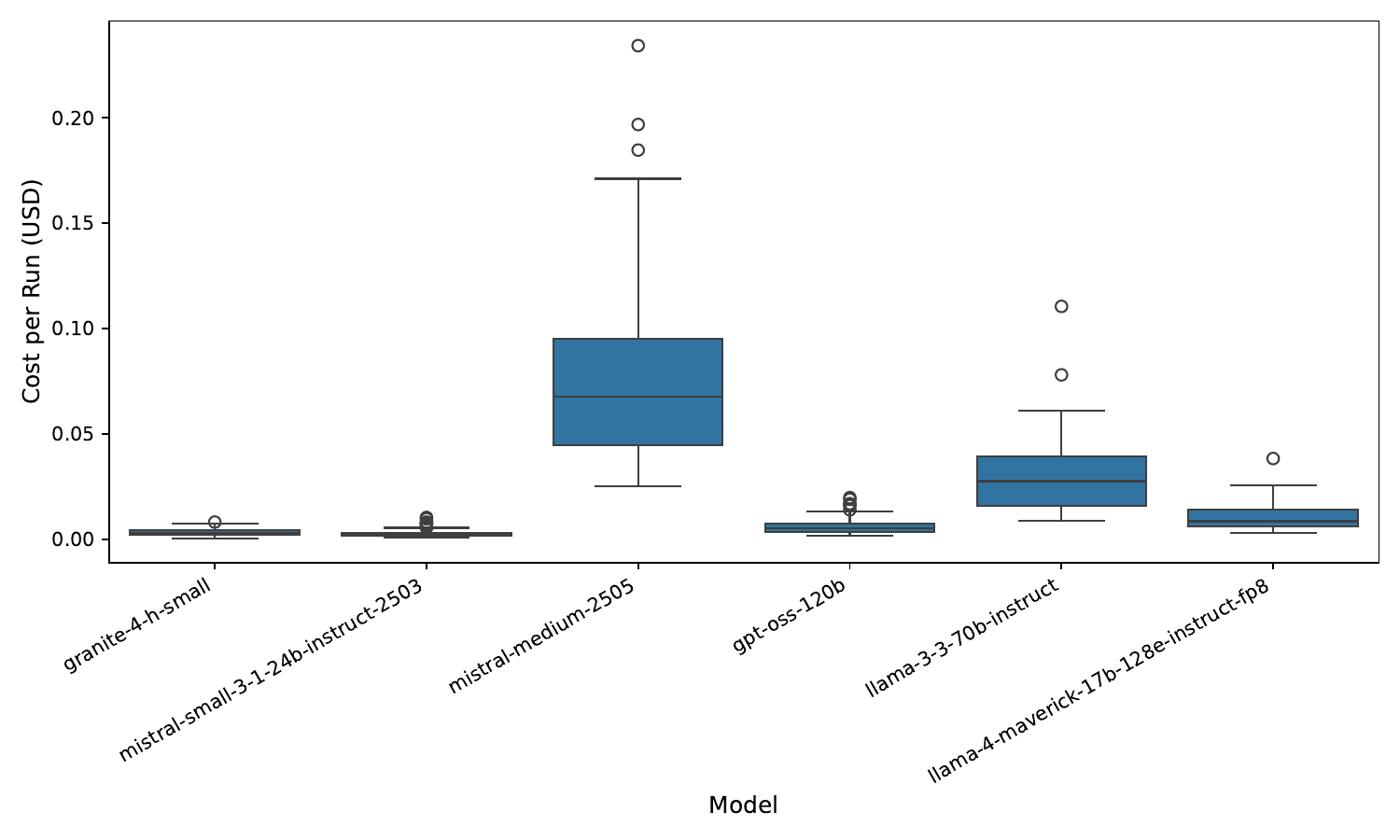}
    \caption{v1}
\end{subfigure}
\hfill
\begin{subfigure}[t]{0.49\textwidth}
    \centering
    \includegraphics[width=\textwidth]{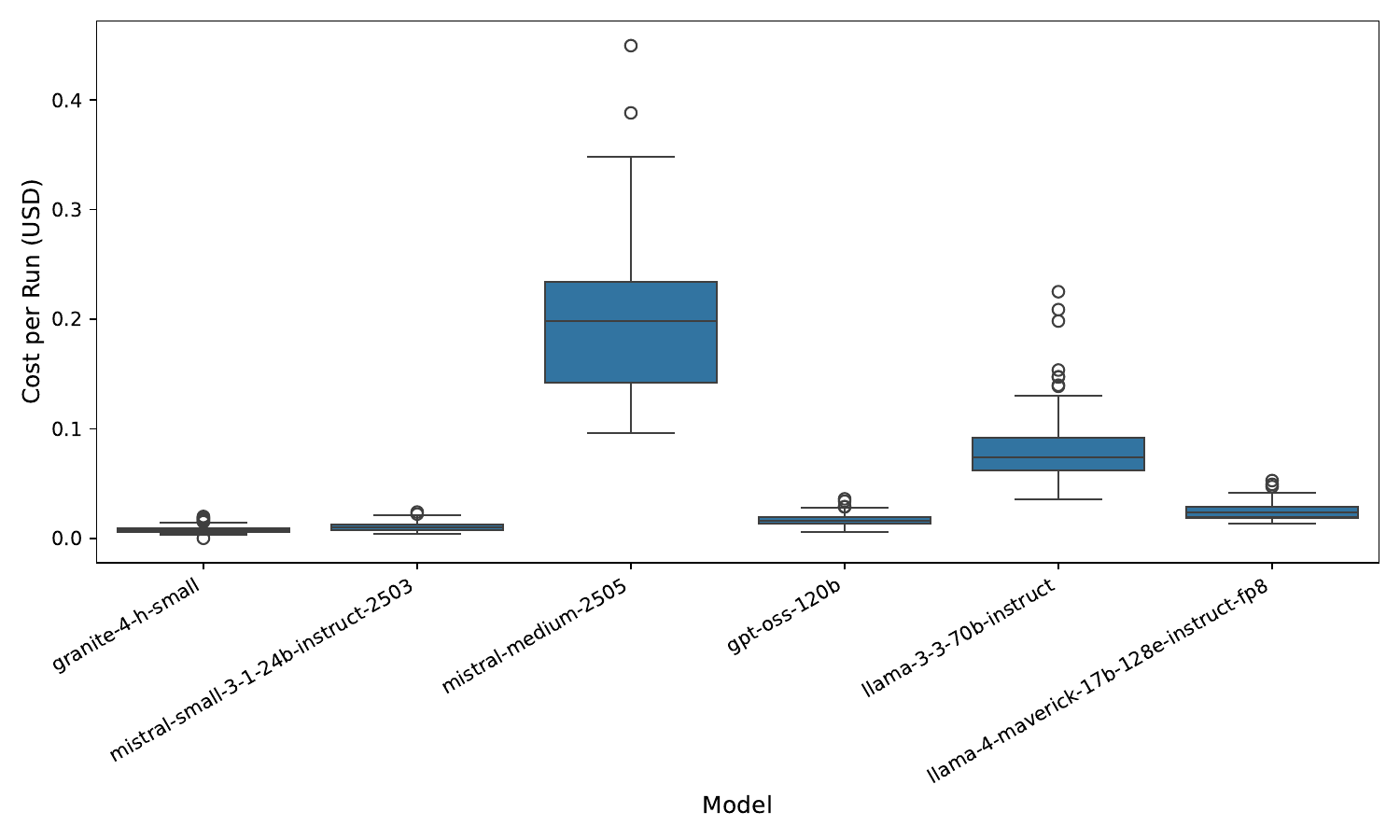}
    \caption{v2}
\end{subfigure}
\caption{Cost distribution per model.}
\label{fig:cost}
\end{figure*}

Figures~\ref{fig:scatter}--\ref{fig:cost} illustrate the success/latency and cost
distributions. The v2 pipeline places four models in the production-viable upper-left
quadrant (${>}$90\% success, ${<}$60s), compared to zero under v1. The approximately linear relationship between input tokens and generation time
(Figure~\ref{fig:tokens}) explains most of the v2 latency increase. The gpt-oss-120b model's
latency improvement despite higher token counts is a notable exception. Structural
success rates measure schema compliance and renderability but do not measure whether
workflows logically fulfill user intent. Section~\ref{sec:semantic} introduces a
complementary semantic evaluation that notably reverses one model's
production-readiness assessment.

\begin{table*}[!t]
\centering
\caption{Model Performance Comparison v1}
\label{tab:v210}
\resizebox{\textwidth}{!}{%
\begin{tabular}{@{}lcccccccc@{}}
\toprule
\textbf{Model} & \textbf{Total} & \textbf{Workflows} & \textbf{Generation} & \textbf{Structural} & \textbf{Structural Success} & \textbf{Avg Time} & \textbf{Avg Cost} & \textbf{Avg Input} \\
 & \textbf{Runs} & \textbf{Generated} & \textbf{Rate} & \textbf{Successes} & \textbf{Rate} & \textbf{(seconds)} & \textbf{(USD)} & \textbf{Tokens} \\
\midrule
granite-4-h-small & 232 & 140 & 60.3\% & 73 & 31.5\% & 41.67 & \$0.0032 & 47,371 \\
mistral-small-3-1-24b-instruct-2503 & 232 & 83 & 35.8\% & 77 & 33.2\% & 34.36 & \$0.0029 & 23,914 \\
mistral-medium-2505 & 232 & 129 & 55.6\% & 91 & 39.2\% & 31.01 & \$0.0745 & 21,273 \\
gpt-oss-120b & 232 & 225 & 97.0\% & 192 & 82.8\% & 99.28 & \$0.0060 & 30,323 \\
llama-3-3-70b-instruct & 232 & 221 & 95.3\% & 123 & 53.0\% & 76.15 & \$0.0290 & 39,405 \\
llama-4-maverick-17b-128e-instruct-fp8 & 232 & 215 & 92.7\% & 134 & 57.8\% & 18.54 & \$0.0102 & 25,191 \\
\bottomrule
\end{tabular}%
}
\end{table*}

\begin{table*}[!t]
\centering
\caption{Model Performance Comparison v2}
\label{tab:v230}
\resizebox{\textwidth}{!}{%
\begin{tabular}{@{}lcccccccc@{}}
\toprule
\textbf{Model} & \textbf{Total} & \textbf{Workflows} & \textbf{Generation} & \textbf{Structural} & \textbf{Structural Success} & \textbf{Avg Time} & \textbf{Avg Cost} & \textbf{Avg Input} \\
 & \textbf{Runs} & \textbf{Generated} & \textbf{Rate} & \textbf{Successes} & \textbf{Rate} & \textbf{(seconds)} & \textbf{(USD)} & \textbf{Tokens} \\
\midrule
granite-4-h-small & 232 & 224 & 96.6\% & 195 & 84.1\% & 52.45 & \$0.0080 & 117,594 \\
mistral-small-3-1-24b-instruct-2503 & 232 & 229 & 98.7\% & 222 & 95.7\% & 42.62 & \$0.0102 & 94,505 \\
mistral-medium-2505 & 232 & 232 & 100.0\% & 223 & 96.1\% & 46.73 & \$0.1969 & 60,507 \\
gpt-oss-120b & 232 & 232 & 100.0\% & 227 & 97.8\% & 66.22 & \$0.0163 & 81,391 \\
llama-3-3-70b-instruct & 232 & 189 & 81.5\% & 172 & 74.1\% & 151.39 & \$0.0785 & 108,447 \\
llama-4-maverick-17b-128e-instruct-fp8 & 232 & 231 & 99.6\% & 210 & 90.5\% & 33.57 & \$0.0244 & 61,487 \\
\bottomrule
\end{tabular}%
}
\end{table*}

\begin{table*}[!t]
\centering
\caption{Performance Improvement: v1 $\rightarrow$ v2}
\label{tab:improvement}
\small
\begin{tabular}{@{}lcccc@{}}
\toprule
\textbf{Model} & \textbf{Generation Rate} & \textbf{Structural Success Rate} & \textbf{Avg Time (s)} & \textbf{Avg Cost (\$)} \\
 & \textbf{v2 / $\Delta$} & \textbf{v2 / $\Delta$} & \textbf{v2 / $\Delta$} & \textbf{v2 / $\Delta$} \\
\midrule
granite-4-h-small & 96.6\% / +36.3\% & 84.1\% / +52.6\% & 52.45 / +10.78 & \$0.0080 / +\$0.0048 \\
mistral-small-3-1-24b-instruct-2503 & 98.7\% / +62.9\% & 95.7\% / +62.5\% & 42.62 / +8.26 & \$0.0102 / +\$0.0073 \\
mistral-medium-2505 & 100.0\% / +44.4\% & 96.1\% / +56.9\% & 46.73 / +15.72 & \$0.1969 / +\$0.1224 \\
gpt-oss-120b & 100.0\% / +3.0\% & 97.8\% / +15.0\% & 66.22 / $-$33.06 & \$0.0163 / +\$0.0103 \\
llama-3-3-70b-instruct & 81.5\% / $-$13.8\% & 74.1\% / +21.1\% & 151.39 / +75.24 & \$0.0785 / +\$0.0495 \\
llama-4-maverick-17b-128e-instruct-fp8 & 99.6\% / +6.9\% & 90.5\% / +32.7\% & 33.57 / +15.03 & \$0.0244 / +\$0.0142 \\
\bottomrule
\end{tabular}
\end{table*}

\section{Semantic Accuracy Evaluation}
\label{sec:semantic}

The structural success rate measured in the preceding sections captures whether a
generated workflow conforms to the Concert Workflows schema and renders correctly in
the UI, but does not measure whether the workflow's chosen action blocks and control
flow fulfill the user's actual request. To complement the structural metric, we
introduced a second evaluation dimension---semantic accuracy---measured by having a
separate large language model act as an external judge of each generated workflow
against the user's original natural-language prompt.

\subsection{Judge Configuration}

The judge was Anthropic Claude Sonnet 4.5, accessed via the IBM BOB Shell at
temperature zero to maximize reproducibility. The judge is independent of all six
models in the benchmark suite (the Granite, Mistral, Llama, and GPT-OSS families),
eliminating any concern about self-favoring bias. The judge scored each workflow
against a structured rubric covering three semantic criteria---completeness,
action-block selection, and logical correctness---and seven platform-convention
criteria covering naming, descriptions, credential handling, and similar conventions,
producing both a 0--100 composite score and a binary ``prompt satisfied'' verdict per
workflow.

\subsection{Protocol}

For each of the 348 (2 pipeline versions $\times$ 6 models $\times$ 29 prompts)
cells in the benchmark, the judge graded three independent workflow generation runs
per cell, yielding 1,044 total gradings. Grading three runs per cell rather than one
allowed us to measure judge stability and to report verdicts as a majority vote
across the three gradings, partially accounting for the stochastic variance between
independent judging runs. Only structurally valid workflows were graded, on the
principle that a user generating workflows with AI would discard invalid results and
retry the pipeline rather than execute a malformed workflow. The three workflows selected for semantic accuracy grading were done so from the eight available structural-validation runs by
retaining outputs whose size met a corruption-detection threshold (at least 40\% of
the largest valid run for that model/prompt/pipeline version). If more than three runs qualified, the lowest-numbered three (by order of generation) were graded.

\subsection{Results}

Semantic accuracy varied substantially across models and pipeline versions, with v2
producing higher mean semantic scores than v1 for five of the six benchmarked models
(Table~\ref{tab:semantic}). The strongest performer overall was
llama-4-maverick-17b-128e-instruct-fp8 on the v2 pipeline, achieving a mean semantic
score of 71.5 and a majority-vote prompt-satisfaction rate of 55.2\%. Even the
strongest model, however, failed the majority-vote satisfaction test on nearly half
of the prompts, indicating that the high structural success rates reported earlier
should not be interpreted as semantic correctness.

\begin{table*}[!t]
\centering
\caption{Per-model semantic accuracy by pipeline version. Mean composite score
(0--100) is reported with its standard error across the 29 prompts per cell.
Satisfaction percentage is the share of prompts for which at least two of the three
grading passes marked the workflow as satisfying the user's request. Rows are sorted
by v2 mean score descending.}
\label{tab:semantic}
\small
\begin{tabular}{@{}lccccc@{}}
\toprule
\textbf{Model} & \textbf{v1 mean (SE)} & \textbf{v1 sat. \%} & \textbf{v2 mean (SE)} & \textbf{v2 sat. \%} & \textbf{Mean delta} \\
\midrule
llama-4-maverick-17b      & 59.8 (2.3) & 27.6\% & 71.5 (2.3) & 55.2\% & +11.7 \\
mistral-small-3-1-24b     & 46.6 (4.6) & 17.2\% & 66.4 (2.5) & 34.5\% & +19.8 \\
llama-3-3-70b-instruct    & 57.2 (2.0) &  6.9\% & 63.6 (3.3) & 34.5\% &  +6.4 \\
mistral-medium-2505       & 49.9 (4.2) & 27.6\% & 61.5 (4.3) & 44.8\% & +11.6 \\
granite-4-h-small         & 57.3 (4.0) & 27.6\% & 58.8 (2.4) & 17.2\% &  +1.5 \\
gpt-oss-120b              & 58.9 (2.9) & 24.1\% & 44.9 (2.1) &  6.9\% & $-$14.0 \\
\bottomrule
\end{tabular}
\end{table*}

\subsection{Key Findings}

The first finding is that semantic evaluation broadly reinforces the v2 pipeline
recommendation made on structural grounds, while quantifying the magnitude of
improvement. Four of the six models showed mean semantic-score gains of six points
or more from v1 to v2, with mistral-small-3-1-24b improving by 19.8 points and
llama-4-maverick-17b by 11.7. These shifts exceed the per-cell standard errors by
wide margins and corroborate the structural finding from a fully independent angle.
The piecewise generation strategy in v2 therefore produces workflows that are not
only more likely to parse, but also more likely to do what the user actually asked
for.

The second finding is that gpt-oss-120b regressed sharply on v2 in a way that only
semantic evaluation could detect. Its mean semantic score fell from 58.9 on v1 to
44.9 on v2, and its majority-vote satisfaction rate dropped to 6.9\%---the lowest
result of any cell in the study---despite gpt-oss-120b achieving the highest
structural success rate (97.8\%) of any model on either pipeline. Inspection of the
judge's reasoning revealed the cause: on v2, gpt-oss-120b frequently produced
``stub'' workflows consisting of a single comment block declaring that the required
actions were unavailable, rather than attempting the integration. These stubs pass
structural validation but are non-executable. The same pipeline change that helped
most models broke this one, which is a useful counter-example to the assumption that
pipeline upgrades benefit all model families uniformly, and a clear illustration of
why structural metrics alone are insufficient.

The third finding is that no benchmarked model achieved reliable correctness, even
on the v2 pipeline. The strongest result observed was 55.2\% majority-vote
satisfaction by llama-4-maverick-17b on v2; the next-best model satisfied only 44.8\%
of prompts under the same metric. This implies that human review of generated
workflows before execution remains necessary, consistent with how the Concert
Workflows platform operates today. A potential extension of the semantic-evaluation
methodology is to surface the judge itself as a post-generation validation check
inside the pipeline: each candidate workflow would be required to pass an LLM judge
that provides iterative feedback before being returned to the user, trading
additional token cost for improved semantic quality. We leave this design direction
to future work.

\section{Analysis}

This section provides a detailed qualitative analysis of each model's performance
characteristics, failure modes, and practical recommendations based on the
experimental results from both pipeline versions.

\subsection{granite-4-h-small}
Piecewise decomposition converted this MoE model's sparse activation (9B active of
32B) from a liability into an asset: each discrete generation stage falls within its
active-parameter budget, whereas monolithic generation exceeded it. Cost remains the
lowest of all models at \$0.0080/workflow. Semantic accuracy improved only marginally
(+1.5 mean score) and majority-vote satisfaction dropped from 27.6\% to 17.2\%,
indicating that structural gains did not consistently translate into better task
fulfillment; piecewise generation helped this model produce valid JSON but not
necessarily the right automation logic.

\textbf{Recommendation}: Suitable for cost-sensitive or GPU-constrained deployments.
Semantic results indicate a human-in-the-loop review pattern is advisable.

\subsection{llama-3-3-70b-instruct}
As a fully dense 70B model, every token in the substantially expanded v2 context
must traverse the full parameter set, producing the worst latency profile in the
study. The generation-rate decline (95.3\% $\to$ 81.5\%) is unexplained by the
architectural change alone and warrants further investigation. Semantic accuracy
improved modestly (+6.4 mean score) but does not compensate for the cost and latency
regression.

\textbf{Recommendation}: Not recommended for production on v2. Strictly dominated by
other options at every accuracy tier.

\subsection{llama-4-maverick-17b-128e-instruct-fp8}
This MoE model (400B total, 17B active, 128 experts) combines large-model reasoning
with sparse-activation efficiency, producing the fastest generation time in v2
(33.57s) while achieving 90.5\% structural success. Critically, it also leads on
semantic accuracy (71.5 mean score, 55.2\% majority-vote satisfaction)---meaning its
workflows are not only structurally valid but more likely to fulfill user intent than
any competitor's.

\textbf{Recommendation}: Strongly recommended for production, especially
latency-sensitive interactive use cases. Recommended default for general-purpose
workflow generation.

\subsection{mistral-small-3-1-24b-instruct-2503}
The largest generation-rate transformation in the study (+62.9pp) signals that v2's
expanded retrieval context unlocks this 24B dense model's capacity in a way v1 did
not. Semantic improvement was also the largest of any model (+19.8 mean score),
though absolute satisfaction (34.5\%) trails llama-4-maverick. At \$0.0101/workflow
it is roughly 2.4$\times$ cheaper than llama-4-maverick at slightly higher structural
accuracy but slower generation.

\textbf{Recommendation}: Strongly recommended as the primary model for cost-sensitive
deployments. For semantically demanding workflows, prefer llama-4-maverick.

\subsection{mistral-medium-2505}
Achieves near-ceiling structural reliability (96.1\%) at 19$\times$ the cost of
mistral-small for a 0.4 percentage-point structural advantage. Semantic improvement
was solid (+11.6 mean score, 44.8\% satisfaction). The cost premium is driven by
Mistral's pricing structure (\$0.003/\$0.01 per 1K input/output tokens), an order
of magnitude above alternatives.

\textbf{Recommendation}: Reserve for mission-critical workflows where structural
failure is unacceptable, or as a fallback. Not recommended for general production
use.

\subsection{gpt-oss-120b}
Leads on every structural metric (97.8\% success, only model with improved latency)
yet ranks last on semantic accuracy (44.9 mean score, 6.9\% satisfaction) on
v2---a sharp regression from v1 ($-$14.0 mean delta). The cause, identified through
judge reasoning, is a failure mode where v2 causes this model to produce structurally
valid but non-executable ``stub'' workflows declaring integrations unavailable rather
than attempting them. This makes gpt-oss-120b uniquely treacherous: first on
structural metrics, last on whether outputs do what users asked.

\textbf{Recommendation}: Not recommended for production on v2 in current
configuration despite leading structural metrics. Prompt engineering to discourage
stub generation is the clear next investigation target.

\section{Conclusion}

Concert Workflows extends prior LLM workflow-generation work to low-code IT
automation at production scale, where cost, latency, and schema compliance impose
additional constraints absent from most academic benchmarks. We analyze how piecewise
decomposition---splitting workflow construction into variable scaffolding, base block
assembly, and nested block generation---enables cost-efficient smaller models (e.g.,
mistral-small at 95.7\% structural success and \$0.01/workflow) to reach production
viability, removing dependency on expensive frontier models.

Semantic evaluation reveals that structural success is necessary but not sufficient.
The gpt-oss-120b result---first on every structural metric yet last on semantic
accuracy---demonstrates a failure mode that structural validation cannot detect:
schema-valid but non-executable stub workflows. This divergence is a general warning
for any production system that uses schema compliance as a proxy for correctness. No
model exceeded 55.2\% majority-vote task satisfaction, confirming that human review
before execution remains necessary. The 8-run aggregation methodology introduced
here, compared to the prior single-run pilot, surfaces this gap more reliably and
provides a replicable baseline for tracking progress as the pipeline and model
landscape evolve. The 29 benchmark prompts are drawn from real-world compliance automation
scenarios---RHSA patching, security group remediation, and configuration drift detection---making production-viable LLM workflow generation directly applicable to
accelerating enterprise compliance programs that operate under tight regulatory SLAs.

\section{Future Work}

The highest priority is strengthening semantic accuracy evaluation. We are developing
a gold-standard benchmark with expert-annotated correct workflows per prompt to
validate judge verdicts against human ground truth and track end-to-end quality over
time. We are also evaluating whether surfacing the judge as a post-generation
validation step inside the pipeline---providing iterative feedback to the
generator---improves end-user outcomes at acceptable cost.

A second priority is per-cell variance characterization. This evaluation aggregates
eight independent runs per cell, but we do not yet report confidence intervals or
significance tests. Follow-up work will expand runs where useful, compute per-prompt
variance to identify unstable generation, and characterize the relationship between
prompt complexity and variance to inform production retry strategies.

We also plan to expand benchmark coverage beyond the compliance domain that anchors
this study, extending prompt classification across the full range of Concert
Workflows automation domains: Security, Networking, Compute, DevOps, Observability,
Resilience, and Risk. This will validate whether the piecewise generation gains
observed here generalize across automation categories with structurally different
action block patterns and control-flow requirements.

Finally, the LLM landscape continues to evolve rapidly. We will maintain a regular
cadence of model evaluation against our benchmark suite to ensure Concert Workflows
AI leverages the best available options.

\section*{Acknowledgments}

The authors used Claude (Anthropic) as an AI writing assistant for proofreading and editorial shortening of this manuscript. AI assistance was applied to the Abstract, Introduction, Related Work, Conclusion, and Future Work sections for grammar validation and length reduction. All evaluation methodology, experimental design, data collection, numerical results, and analytical conclusions are entirely the work of the authors; no AI system was used to generate or interpret any experimental content.

\bibliographystyle{IEEEtran}
\bibliography{references}

\end{document}